\documentclass[runningheads]{llncs}
\usepackage{graphicx}
\usepackage{booktabs} 

\usepackage{amsmath}
\usepackage{balance}       
\usepackage{graphics}      
\usepackage{color}
\usepackage{enumitem}
\usepackage{algorithm}  
\usepackage{algorithmic}  
\usepackage{CJKutf8}
\usepackage{multirow}
\usepackage{dblfloatfix}
\usepackage[sort]{cite}
\usepackage{capt-of}    
\usepackage{adjustbox}

\usepackage{gb4e}
\noautomath
\usepackage{lipsum}
\usepackage{cuted}

\usepackage{todonotes}

\definecolor{REd}{rgb}{0.5,0.2,0.2}
\definecolor{GREEn}{rgb}{0.2,0.4,0.2}
\definecolor{BLUe}{rgb}{0.2,0.2,0.6}

\DeclareMathOperator*{\argmax}{\Huge arg\,max}

\begin{document}
\title{Adaptive Learning Material Recommendation in Online Language Education}
\titlerunning{Adaptive Learning Material Recommendation in Online Language Education}
%
\author{Shuhan Wang\inst{1} \and Hao Wu\inst{2} \and Ji Hun Kim\inst{1} \and Erik Andersen\inst{1}}
\authorrunning{Wang et al.}
%
\institute{Department of Computer Science, Cornell University \and
Department of Computer Science, George Washington University\\
\email{\{sw788, jk2227, ela63\}@cornell.edu, fqq11679@gmail.com}}
\maketitle              

\begin{abstract}


Recommending personalized learning materials for online language learning is challenging because we typically lack data about the student's ability and the relative difficulty of learning materials. This makes it hard to recommend appropriate content that matches the student's prior knowledge. In this paper, we propose a refined hierarchical knowledge structure to model vocabulary knowledge, which enables us to automatically organize the authentic and up-to-date learning materials collected from the internet. Based on this knowledge structure, we then introduce a hybrid approach to recommend learning materials that adapts to a student's language level. We evaluate our work with an online Japanese learning tool and the results suggest adding adaptivity into material recommendation significantly increases student engagement. 

\end{abstract}

\section{Introduction}


Keeping students engaged with personalized content in online language learning presents challenges in both the selection of learning content and assessment of students' abilities. The content-side challenge is how to prepare a well-organized corpus of learning materials that are labeled with their difficulty. This is especially hard for online language learning systems that leverage a large amount of up-to-date learning materials collected from the Internet, since it is prohibitively expensive to ask experts to measure the difficulty of those materials. Furthermore, existing data-driven approaches of automatic difficulty evaluation fail because of the lack of student data for up-to-date online content. The student-side challenge is how to assess a student's competency level and recommend content that is appropriate for the prior knowledge of that student. Most existing content recommender systems for language learning are designed for formal learning scenarios such as universities and schools, and they make recommendations based on the student's standardized pre-assessment results. However, these systems cannot be scaled to informal learning scenarios such as online learning, where we usually do not have accurate and standardized information of a student's prior knowledge. Moreover, existing assessment and recommendation systems~\cite{lord1980applications,rasch1993probabilistic,chen2006personalized} usually use unidimensional measurements for student ability and content difficulty, which is not comprehensive~\cite{falmagne2006assessment}. Ideally, we would have a unified system that can multidimensionally evaluate a student's ability and the relative difficulty of learning materials in order to prepare future lessons for that student, without requiring the student's prior information or significant expert labor. 

Previous work on multidimensional knowledge structuring for grammar knowledge uses strict constraints to specify the relative difficulty between two texts~\cite{wang2017unified}. However, this does not scale to teaching vocabulary with a large online corpus since these strict constraints yield too few edges in the structure. To this end, we investigated how to increase density without suffering an unacceptable loss of quality in prediction of relative difficulty. We propose the \emph{fuzzy partial ordering graph}, a refined hierarchical knowledge structure with relaxed constraints. 

In this paper, we present a material recommender system for online language learning that incorporates adaptive knowledge assessment. It collects authentic and up-to-date learning materials from the Internet and organizes them with a fuzzy partial ordering graph. It also uses a probabilistic function to balance assessment and recommendation throughout the learning process in order to improve student engagement\footnote{Students often quit quickly while using online learning tools~\cite{butler2015automatic}. Therefore, our main focus is increasing engagement and time on task as opposed to learning efficiency.}. To evaluate our fuzzy partial ordering graph and adaptive recommendation approach, we developed \emph{JRec}, an online Japanese language learning tool that aims to recommend appropriate reading texts from the Internet based on the student's prior knowledge. A user study of \emph{JRec} demonstrates that our adaptive recommendation system led users to read 62.5\% more texts than a non-adaptive recommendation version. This result indicates that the fuzzy partial ordering graph successfully enables a multidimensional assessment of the student's vocabulary knowledge, which can be incorporated in our adaptive recommendation algorithm in order to improve engagement.

\section{Related Work}

\subsection{Hierarchical Knowledge Organization}

Andersen et al. introduced a technique for automatic knowledge organization on procedural tasks~\cite{andersen2013trace}. This technique characterizes each task by analyzing the \emph{execution trace} of solving it, and studies the \emph{partial orderings} between task-solving procedures to build the knowledge structure in a specific domain. More recently, Wang et al. applied partial orderings to build hierarchical knowledge graphs in non-procedural domains, such as natural language grammar~\cite{wang2017unified}. This model takes advantage of compositionality, the idea that a practice problem can be described as a multiset of conceptual units~\cite{wang-andersen:2016:COLING}. Within the partial ordering graph, problem $a$ is easier than problem $b$ (indicated as an edge from $a$ to $b$) if $b$ covers all conceptual units of $a$. However, this model cannot be applied to educational domains with a large number of conceptual units, such as vocabulary learning, since partial ordering graphs in those domains are too sparse to use~\cite{wang2017unified}. We build on this work by relaxing the relationship between practice problems and introducing the \emph{fuzzy partial ordering graph} to ensure that the hierarchical structure of vocabulary knowledge is sufficiently dense.

\subsection{Knowledge Assessment and Computer-based Test}

Item Response Theory~(IRT) provides a well-established framework for knowledge assessment~\cite{drasgow1990item,hambleton1991fundamentals,reckase2009multidimensional,embretson2013item}. IRT stipulates that a student's response to an item is a function of student ability and item parameters~(primarily, item difficulty)~\cite{lord1980applications,rasch1993probabilistic}. IRT is also a crucial tool in Computerized Adaptive Testing~(CAT)~\cite{weiss1984application,wainer1990item,van2000computerized,yao2012multidimensional}. CAT uses IRT to select the items that can best discriminate examinees and updates the estimate of exaiminee abilities according to their responses. Both IRT and CAT characterize an item by statistically analyzing large amounts of student responses. However, this does not apply to the fresh materials in online learning due to the lack of sufficient student data. Moreover, most IRT and CAT approaches use unidimensional measurements for item difficulty (or a fixed number of pre-defined dimensions), which is incomprehensive~\cite{falmagne2006assessment}. In contrast, our work measures the difficulty of online learning materials by studying the compositionality of domain knowledge and building the hierarchical knowledge structure within the corpus. By doing this, our system is able to leverage fresh learning materials from the Internet, and make appropriate recommendations for each student.

\subsection{Educational Recommender Systems}
Researchers have developed many Educational Recommender Systems~(ERS) based on students' prior knowledge~\cite{chen2006personalized}, topics of interest~\cite{hsu2010development} and learning styles (e.g. verbal/visual, active/reflective)~\cite{latham2014adaptation,hwang2013learning},  However, most of these ERS systems are designed for formal learning scenarios, such as learning in universities. In formal learning, materials are measured and organized with well-defined structure or metadata by experts~\cite{latham2014adaptation}, and students are characterized with standard pre-assessments~(for prior knowledge)~\cite{chen2006personalized} or pre-questionnaires~(for learning preferences such as topics of interest and learning style)~\cite{hsu2010development,hwang2013learning}. However, in informal scenarios such as online learning, a huge amount of learning materials cannot be manually structured and indexed with domain concepts and metadata (the `open corpus problem')~\cite{brusilovsky2007open}, and the modeling of students is either lacking or unstandardized~\cite{manouselis2011recommender}. In this paper, we aim to address these issues in online learning. Our recommender system automatically organizes the learning content from the Internet into a hierarchical model and incorporates adaptive assessment into the recommender system in order to improve student engagement.

\section{Modeling Vocabulary Knowledge}

The Internet provides a vast corpus of reading materials that are suitable for language learning. However, due to the large size and the freshness of this corpus, it is prohibitively expensive to ask experts to measure the difficulty of those materials, and data-driven techniques do not apply either due to the lack of student data. Therefore, in order to leverage learning materials from the Internet, recommender systems should be able to automatically measure the difficulty of those materials and build the hierarchical knowledge structure within the corpus. In this section, we first summarize how existing work did this for grammatical knowledge, then discuss an issue with this work that limits its application with regard to vocabulary. Subsequently, we address this issue and propose a refined hierarchical structure for modeling vocabulary knowledge.

\begin{figure}[t]
\centering
\includegraphics[width=0.8\textwidth]{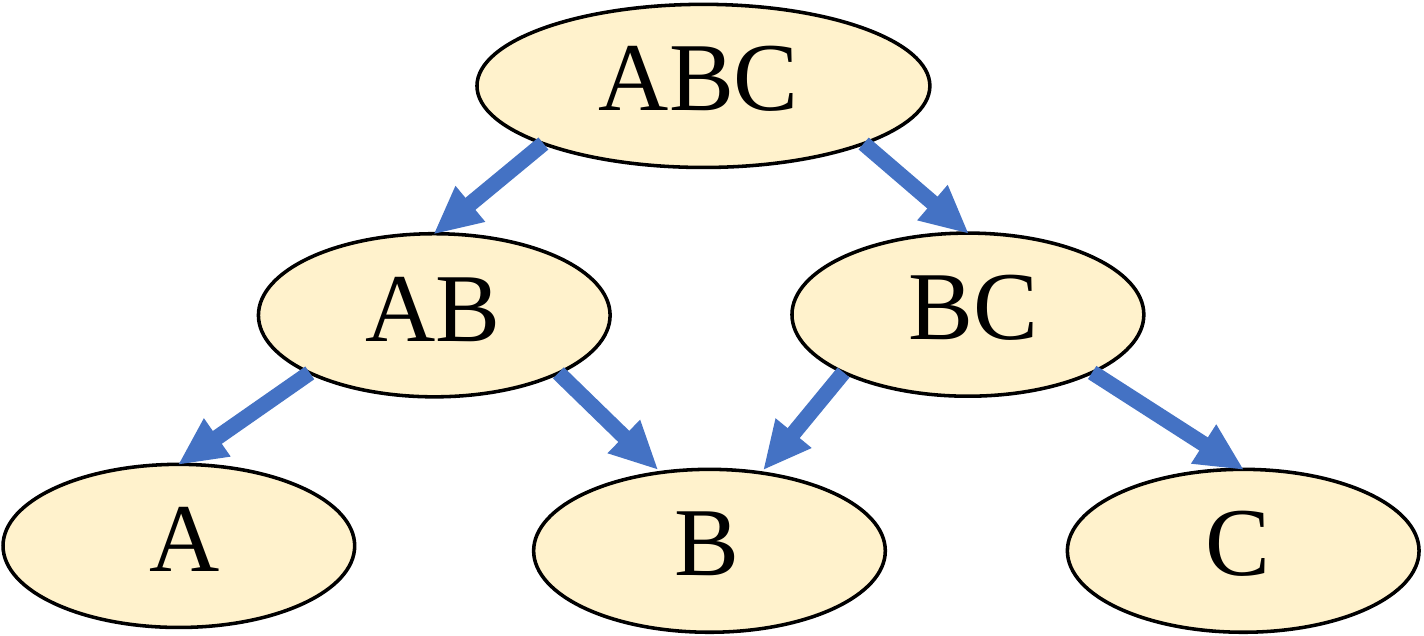}
\caption{A sample partial ordering graph. Each node represents a practice problem containing a specific set of required concepts. Directed edges represent a ``directly harder than'' relation between two problems.}
\label{fig:po_graph}
\end{figure}

Recently, we used partial ordering graphs to model the relationship between reading materials and model the hierarchical structure of grammatical knowledge in a corpus~\cite{wang2017unified}. We briefly recap this previous work here:
\vspace{-1mm}
\begin{itemize}[noitemsep]
\item 
A practice problem (a reading text) can be characterized as a \emph{multiset} of its required concepts.
\item
Problem $s_1$ is \emph{harder than} problem $s_2$~(indicated as $s_1 > s_2$) if and only if $s_1$ covers all required concepts of $s_2$. This also implies that students who understand $s_1$ will also be able to understand $s_2$.
\item 
Problem $s_1$ is \emph{directly harder than} problem $s_2$ if $s_1 > s_2$, and there is no other problem $s_3$ such that $s_1 > s_3 > s_2$.
\item 
A partial ordering graph is a Direct Acyclic Graph~(DAG): each node represents a practice problem and each edge represents a ``directly harder than'' relation between two problems.
\vspace{-1mm}
\end{itemize}

\noindent Figure~\ref{fig:po_graph} shows a sample partial ordering graph. The partial orderings are useful because they can help in the modeling of students' knowledge: a student understanding problem $s$ implies that he/she can also understand problems easier than $s$. Also, this model takes advantage of \emph{compositionality} of practice problems~\cite{wang-andersen:2016:COLING}, and the order of concepts within a problem is unimportant. Therefore, it can be applied to both procedural and non-procedural educational tasks.

However, this work also mentions that in order for the partial ordering graph to work, the hierarchical structure of domain knowledge must be ``sufficiently dense''. Otherwise, the partial ordering graph will only have a small number of edges, and there will not be enough partial ordering relations that can be used. Therefore, this model cannot be directly applied to vocabulary knowledge because vocabulary learning requires a large amount of conceptual units. For example, there are over 10,000 vocabulary words in Japanese learning whereas there are only around 500 grammatical concepts. A typical Japanese sentence may require 10-30 vocabulary words compared to only around 5 grammatical concepts. As a result, it is not common in an authentic corpus that a sentence covers all vocabulary knowledge of another sentence, and the vocabulary-based partial ordering graph will be too sparse. 

\begin{figure}[t]
\centering
\includegraphics[width=0.8\textwidth]{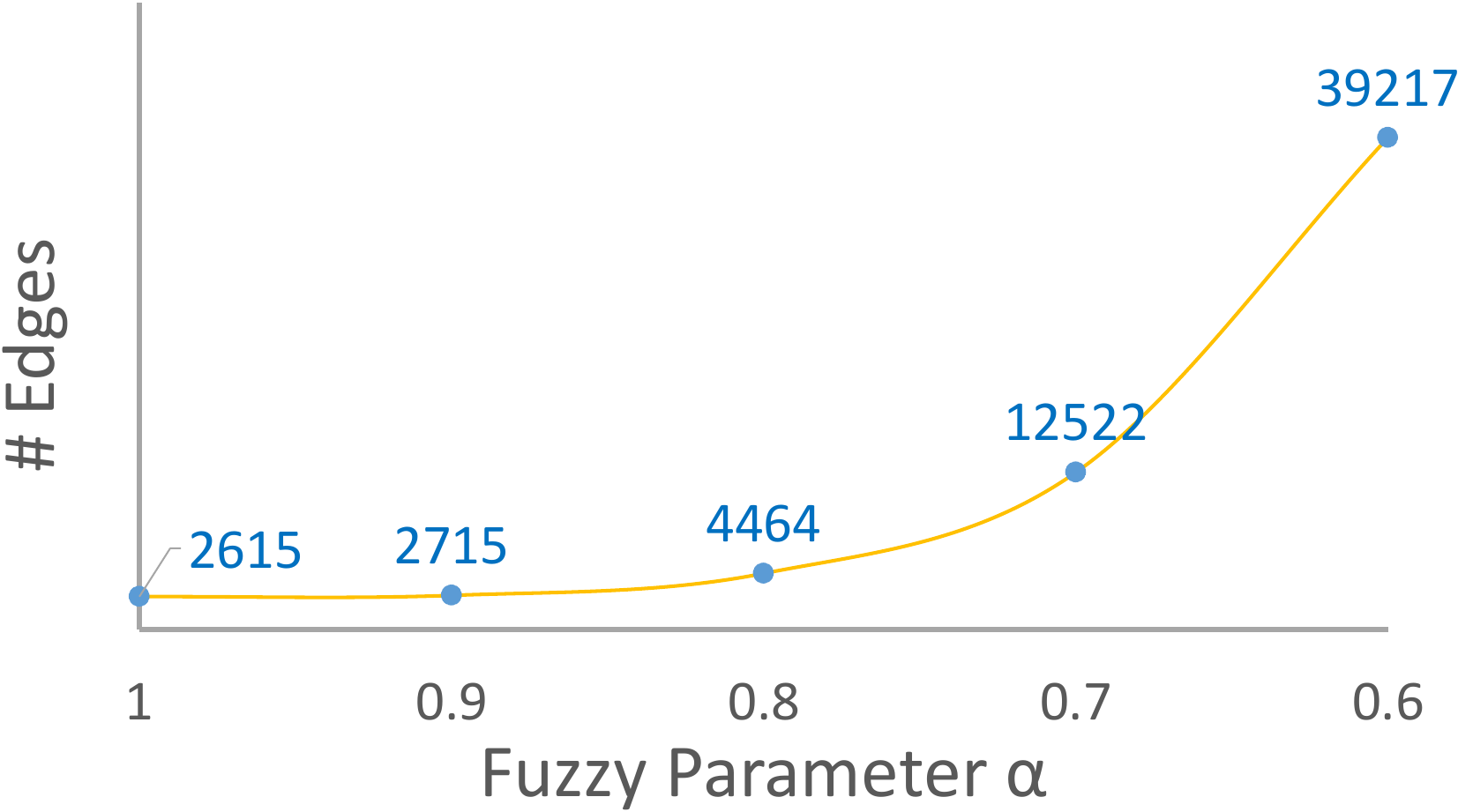}
\caption{Number of edges in the fuzzy partial ordering graph over different fuzzy parameters, in our corpus of 4,269 texts. Decreasing the fuzzy parameter will increase the graph density exponentially.}
\label{fig:fuzzy_density}
\end{figure}

To address this, we take advantage of the idea supported by existing work~\cite{krashen1985input,vygotsky1980mind} that language learners can infer the meanings of some unknown words if they understand the majority of the text, and they will accumulate language knowledge in this way. This idea inspired us to relax the partial ordering relations between two texts in order to increase the density in the vocabulary-based hierarchical knowledge structure.

\vspace{\parskip}
\begin{definition}
Problem $s_1$ is \emph{$\alpha$-fuzzily harder than} problem $s_2$ if $s_1$ covers at least a proportion $\alpha$ of required concepts of $s_2$. Using this fuzzy partial ordering, we can also define the \emph{fuzzy partial ordering graph}.
\label{def:fuzzy}
\end{definition}

We found that the hierarchical knowledge structure based on the fuzzy partial ordering in Definition~\ref{def:fuzzy} has 71\% more edges than the strict version introduced in the former work~\cite{wang2017unified}, using fuzzy parameter $\alpha=0.8$. As the fuzzy parameter $\alpha$ decreases, the number of edges in the fuzzy partial ordering graph increases exponentially~(Figure~\ref{fig:fuzzy_density}). Although this relaxation increases density, it also lowers our confidence in the fuzzy partial ordering relations. If $\alpha$ is too small, there will be many edges in the fuzzy partial ordering graph, but our confidence in each edge (namely, 
the likelihood that a student understands a problem if he/she understands another problem that is fuzzily harder than it) will be too low.

\begin{CJK}{UTF8}{min}
\begin{table*}
\centering
\begin{tabular}{|c|l|}
\hline
$\alpha$ & \multicolumn{1}{c|}{Sample Sentence Pair}\\
\hline
0.9 &
\begin{minipage}[h]{0.95\linewidth}
\scriptsize{
\vspace{0.05in}
\gll 席   の 数     より 客        のほうが 多かった ことは、 ５回      ありました\\ 
     seat of number than passenger {}        more    {}       {5 times} {there was}\\
\vspace{-0.05in}     
\glt \textbf{``There were 5 times when there were more passengers than the number of seats.''}\\
\vspace{-0.05in}
\gll 先月、 全日空 の 飛行機 が、 席 の 数 より 客 が １人 多い まま 出発しよう としたことが ありました\\
    {last month} ANA of {flight} {} seat of number than passenger {} {1 person} {more} {} {about to depart} {one time} {there was}\\
\vspace{-0.05in} 
\glt \textbf{``Last month, there was a time when an ANA flight was about to depart but there was one more passenger than the number of seats.''}\\
\vspace{0.05in}
}
\end{minipage}
\\\hline
0.8 &
\begin{minipage}[h]{0.95\linewidth}
\scriptsize{
\vspace{0.05in}
\gll ９日、 この ボランティアに なり   たい 人たちが 集まって、 太田市    で 勉強しました\\ 
      9th   this volunteer      become want people   gather     Ota(city) in  studied \\
\vspace{-0.05in} 
\glt \textbf{``On the 9th, people who wanted to become volunteers gathered and studied in Ota.''}\\
\vspace{-0.05in}
\gll 集まった 人たち は、 あと ２回      勉強して テストに 合格する と、 病院     など で 通訳をする ボランティア になります\\
     gathered people {}   more {2 times} study    test     pass     if   hospital like in interpret  volunteer    become\\
\vspace{-0.05in} 
\glt \textbf{``The gathered people will become volunteer interpreters in places like hospitals, if they study two more times and pass the test.''}\\
\vspace{0.05in}
}
\end{minipage}
\\\hline

0.7 &
\begin{minipage}[h]{0.95\linewidth}
\scriptsize{
\vspace{0.05in}
\gll シリア で は、 政府       と  政府       に 反対する 人たち の 戦争 が 続いています\\ 
     Syria  in {}   government and government {} against  people of war  {} {is ongoing}\\
\vspace{-0.05in} 
\glt \textbf{``In Syria, the war between the government and the anti-government faction is still ongoing.''}\\
\vspace{-0.05in}
\gll 政府      に 反対する 人たち が たくさん いる       アレッポという町 に は、 政府       の 軍    が ２週間も  空  から 攻撃を 続けています\\
    government {} against  people {} many     {there is} {Aleppo(city)}      in {}   government of army  {} {2 weeks} air from attack maintaining\\
\vspace{-0.05in} 
\glt \textbf{``In Aleppo, where there is a large anti-government faction, the government army maintained attacks from the air for two weeks.''}\\
\vspace{0.05in}
}
\end{minipage}
\\\hline

0.6 &
\begin{minipage}[h]{0.95\linewidth}
\scriptsize{
\vspace{0.05in}
\gll シリア で は、 政府       と  政府       に 反対する 人たち の 戦争 が 続いています\\ 
     Syria  in {}   government and government {} against  people of war  {} {is ongoing}\\
\vspace{-0.05in} 
\glt \textbf{``In Syria, the war between the government and the anti-government faction is still ongoing.''}\\
\vspace{-0.05in}
\gll 今  まで  の １０年、    私 は 戦争 が 続いている    所    や 難民    が 生活している 所    へ 何度も               行きました\\
     now until of  {10 years} I  {} war  {}  {is ongoing} place {} refugee {}  {is living} place to {for multiple times} went\\
\vspace{-0.05in} 
\glt \textbf{``In the last 10 years, I have made multiple visits to places where a war was ongoing or refugees were living.''}\\
\vspace{0.05in}
}
\end{minipage}
\\\hline
\end{tabular}
\caption{Sample Sentence Pairs in the fuzzy partial orderings with the fuzzy parameter $\alpha=$ 0.9/0.8/0.7/0.6. For each fuzzy parameter $\alpha$, the second sentence is $\alpha$-fuzzily harder than the first sentence. As the fuzzy parameter $\alpha$ decreases, our confidence in the fuzzy partial orderings (the likelihood for a student to understand the first sentence if he/she understands the second one) also drops.  Text source: NHK Easy~\cite{website:nhkeasy}.}
\label{tab:fuzzy_pair}
\end{table*}
\end{CJK}

This leads to a trade-off between the density of the hierarchical knowledge structure and our confidence in the (fuzzy) partial ordering relations. To identify the best fuzzy parameter for structuring vocabulary knowledge, we conducted a case study in our corpus of 4,269 Japanese texts. Examples of ``fuzzily harder than'' sentence pairs for fuzzy parameter $\alpha$=0.9/0.8/0.7/0.6 are listed in Table~\ref{tab:fuzzy_pair}. We believe that the $\alpha=0.9$ and $\alpha=0.8$ values are suitable for use. In these two cases, the second sentence covers almost all the vocabulary knowledge in the first sentence. Therefore, students are very likely to understand the second sentence if they understand the first one. However, our confidence in the fuzzy partial ordering relations are too low for the $\alpha=0.7$ and $\alpha=0.6$ values, since in these two cases, the first sentence requires a certain amount of vocabulary knowledge that is not required by the second sentence. In this situation, we cannot be sure students who understand the second sentence will also understand the first one.

Based on these results, we used the fuzzy parameter $\alpha=0.8$ in our vocabulary-based fuzzy partial ordering graph because the graph is sufficiently dense and our confidence in the partial ordering relations are high enough to use. However, the optimal fuzzy parameter $\alpha$ is likely different in each educational domain and needs to be empirically studied in each domain.

\section{Adaptive Learning Material Recommendation}

In order for students to be engaged, they need to experience learning materials at the right difficulty level. Although we have seen existing educational recommender systems that recommend learning materials based on student ability, most of these systems characterize each student by standardized pre-assessment results, such as in standard language placement tests~\cite{chen2006personalized,DBLP:journals/corr/PilanVB17}. However, in online learning, where pre-assessment results are usually unavailable, we still lack an effective approach to recommend learning materials that automatically assesses and adapts to each student's prior knowledge. 
To improve this, we seek to build a recommender system that carefully balances the trade-off between assessment and recommendation: in order for recommendations to be appropriate, the system needs to accurately assess each student; however, excessive assessment can potentially harm engagement because students might need to respond to too many problems that are far outside of their comfort zone. 


\subsection{Adaptive Assessment Heuristic}

To recommend learning materials that adapt to each student's prior knowledge, we follow a typical interaction process in adaptive education systems~\cite{yao2012multidimensional, wang2017unified}: the system keeps selecting the next problem~(learning material) to present to a student and updating the model of the student's knowledge based on his/her response. 
We previously proposed a framework for modeling a student's knowledge in the hierarchical knowledge structure~\cite{wang2017unified}. This framework characterizes a student's knowledge by monitoring whether he/she can solve each problem in the library. With the help of partial orderings between problems, the assessment algorithm can infer the student's performance on some problems without presenting them. To be more specific, if the student can solve problem $s_1$, he/she can also solve problems that are easier than $s_1$; if the student cannot solve problem $s_2$, he/she cannot solve problems that are harder than $s_2$ either. 

Building on this framework, we propose an adaptive assessment heuristic to select the next problem in the (fuzzy) partial ordering graph.

\textbf{The (Adaptive) Assessment Heuristic:} Select the problem that maximizes the \emph{expected} amount of information gained on the student's prior knowledge. Formally, the assessment heuristic selects the problem $s^*$ such that:
\vspace{-1mm}
\begin{equation}
    s^* = \argmax_s{\ [\,p_sn_s^++(1-p_s)n_s^-\,]}
    \label{equ:assessment}
    \vspace{-1mm}
\end{equation}
where $p_s$ indicates the probability that the student can solve $s$. If the student can solve $s$, $n_s^+$ represents how many problems we know that he/she can solve. Otherwise, if the student cannot solve $s$, $n_s^-$ represents how many problems we know that he/she cannot solve. Both $n_s^+$ and $n_s^-$ include $s$ itself and exclude the problems we already know the student can/cannot solve before presenting $s$. 

The probability $p_s$ can be estimated in a straightforward way:
\vspace{-1mm}
\begin{equation}
    p_s = N^+/(N^++N^-)
    \vspace{-1mm}
\end{equation}
where $N^+$ and $N^-$ denote the number of presented problems that the student can and cannot solve.

Note that our assessment heuristic in Equation~(\ref{equ:assessment}) is different from existing work~\cite{wang2017unified} since our heuristic incorporates the probability $p_s$ and calculates the \emph{expected} amount of information gained on the student's prior knowledge, while existing work only focuses on the information gained in the lesser of the two cases where the student can/cannot solve the problem. By doing this, our heuristic adapts to students at the extremes of ability levels much faster than existing work. For instance, for a very good student that can solve 9 out of 10 problems presented to him/her, our assessment heuristic will start to select the hardest problems in our library from the fifth problem, while the heuristic in existing work will always select the problems with intermediate difficulty.

\subsection{ZPD-based Recommendation Heuristic}

Vygotsky's Zone of Proximal Development~(ZPD) stipulates that a student can solve the problems just beyond his/her knowledge with guidance, and a good teacher/tutor system should recommend those problems to the student. Based on this theory, we propose the recommendation heuristic to select the next problem in the (fuzzy) partial ordering graph.

\textbf{The (ZPD-based) Recommendation Heuristic:} Select the problem that is directly harder than some problem that the student can solve. Since we believe that students are more engaged while solving a problem relevant to their experience, if there are multiple problems satisfying this requirement, pick the one that is \emph{most relevant} to the student prior knowledge.

Here the relevance of a problem to the student's prior knowledge can be measured by counting the ``harder than'' relations between that problem to any problem that the student can solve within the hierarchical knowledge structure. Practically, the relevance is measured as the number of edges from that problem's node to any solvable problem's node in the (fuzzy) partial ordering graph.

\subsection{Balancing Assessment and Recommendation}

Both assessment and recommendation heuristics are for selecting the next problem to present to students. The difference between them is that the assessment heuristic searches the whole knowledge structure to extract more information about a student's knowledge, while the recommendation heuristic only selects the problems that are just outside the ``boundary'' of
the set of problems that the student has correctly answered.

Our system uses a probabilistic function to balance the assessment and recommendation heuristics. To select the next problem, our system chooses the assessment heuristic with probability 
    \begin{equation}
        p=\#Prob/M
        \label{equ:balance}
    \end{equation}
and chooses the recommendation heuristic with probability $1-p$. Here \,$\#Prob$ \,represents the number of the problems that the student has experienced, regardless of whether he/she has solved those problems. $M$ is a pre-set parameter that controls how fast our system transitions from assessment-favoring to recommendation-favoring. It also indicates that our system will always choose the recommendation heuristic after the student has experienced $M$ problems. 

This function ensures that our system favors the assessment heuristic at the beginning in order to gain more information about a student's knowledge. As the student experiences more problems, and the model of student's knowledge gets more comprehensive and convincing, our system tends to make more recommendations in the student's ZPD.



\begin{figure}
\centering
\includegraphics[width=0.8\textwidth]{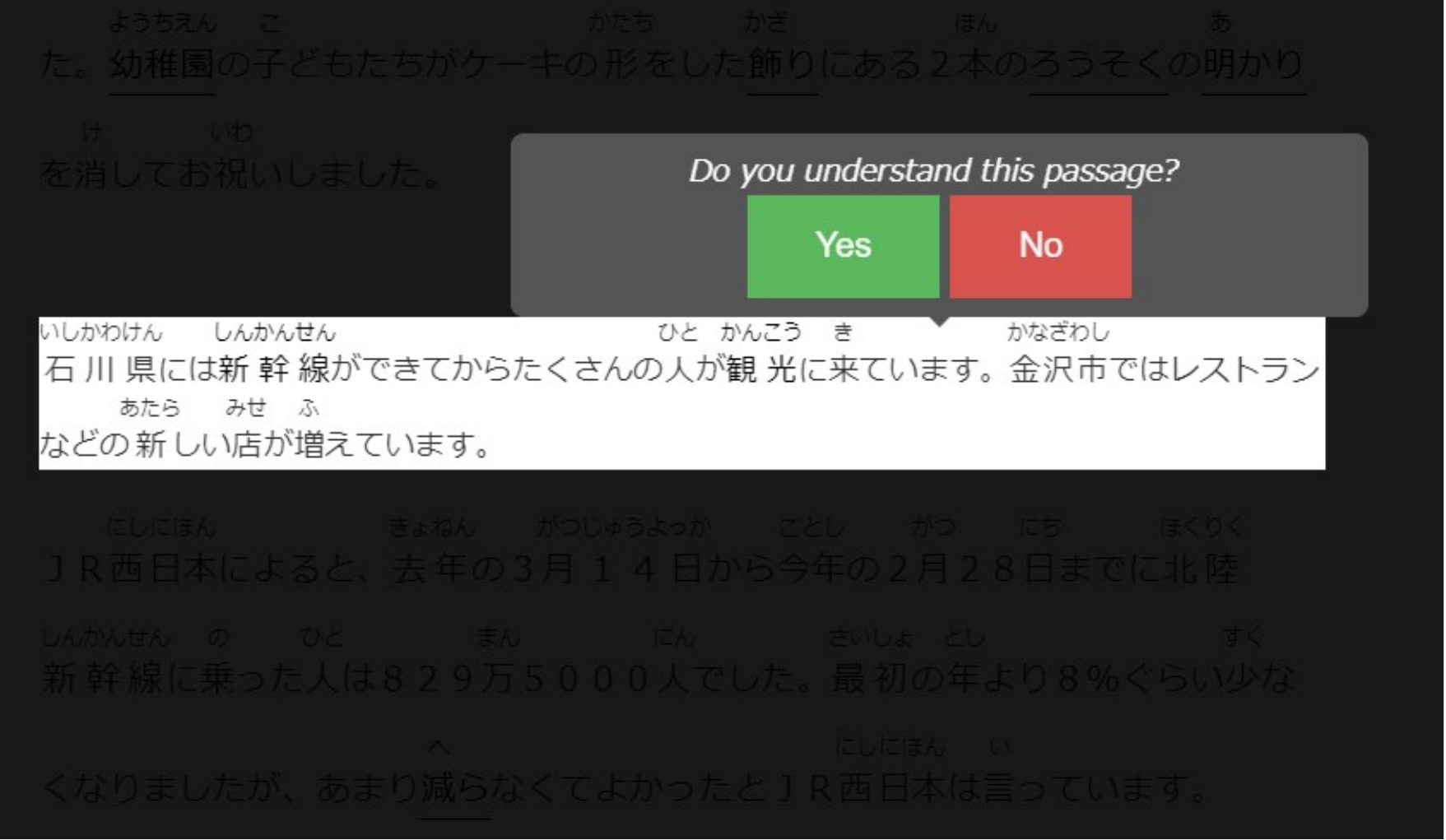}
\caption{Screenshot of \emph{JRec}, a Japanese reading text recommendation tool. It draws texts from NHK Easy~\cite{website:nhkeasy}. When using this tool, users are directed to an NHK Easy webpage, read a recommended text, and respond whether or not they understand it. Our tool highlights the recommended text and grays out the rest of the webpage.}
\label{fig:jrec}
\vspace{0.5in}
\centering
\includegraphics[width=0.8\textwidth]{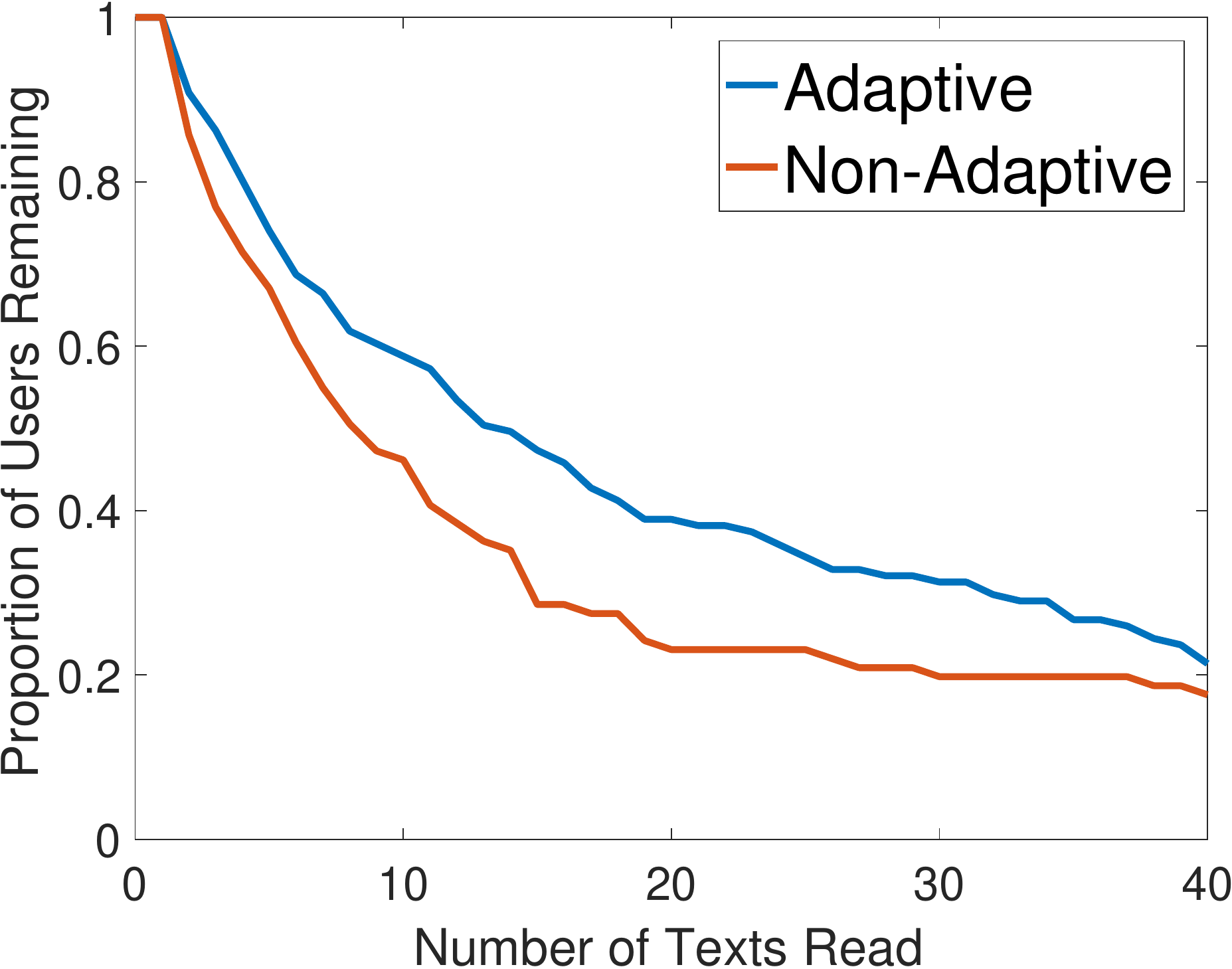}
\caption{Proportion of users remaining after reading certain amount of texts. We observed that the median user in the adaptive recommendation group read 62.5\% more texts than that in the non-adaptive recommendation group, which indicates that incorporating adaptive assessment significantly improved student engagement in learning material recommendation.}
\label{fig:jrec_results}
\end{figure}

\section{Evaluation of Adaptive Recommendation}
We evaluate our adaptive learning material recommender system in \emph{JRec}, a Japanese reading text recommendation tool. Our corpus of 380 articles was collected from NHK Easy~\cite{website:nhkeasy}, a Japanese news website for language learners. In order to accommodate beginners, our tool split those articles into 4,267 sentences and paragraphs so that students do not have to read the whole article. Afterwards, it analyzed the hierarchical structure of vocabulary knowledge in the corpus and built a fuzzy partial ordering graph. When using this tool, users are directed to an NHK Easy webpage, read a recommended text (a paragraph or a sentence), and respond whether or not they understand it. Our tool highlights the recommended text and grays out the rest of the webpage. Figure~\ref{fig:jrec} shows a screenshot of JRec.
We released our tool in the Japanese Learning Sub-reddit~\cite{website:reddit} and recruited 368 users in three days.


\subsection{Adding Adaptivity Improved Engagement Significantly}

In \emph{JRec}, we tested four different versions: 1) adaptive recommendation (which balances recommendation and assessment as we discussed in the last section\footnote{We used $M=50$ in Equation~(\ref{equ:balance}) to balance assessment and recommendation.}) and 2) non-adaptive recommendation (with no assessment incorporated), as well as 3) assessment-only, and 4) random selection as additional baselines. We particularly wanted to see if adaptive recommendation is more engaging than non-adaptive recommendation, since this would demonstrate that incorporating adaptive assessment can enhance learning material recommendation.

\begin{table}[t]
    \renewcommand{\arraystretch}{1.1}
    \centering
    \adjustbox{width=1\textwidth}{
    \begin{tabular}{|r|c|}
    \hline
    Comparison & Results\\
    \hline
    Adaptive Recommendation. vs Non-adaptive Recommendation. & $p = .035, Z = 2.109$\\
    Assessment-Only vs Adaptive Recommendation & $p = .766, Z = 0.298$\\
    Assessment-Only vs Non-adaptive Recommendation & $p = .022, Z = 2.287$\\
    Random vs Non-adaptive Recommendation & $p = .547, Z = 0.603$\\
    Assessment-Only vs Random & $p = .294, Z = 1.049$\\
    Adaptive Recommendation vs Random & $p = .389, Z = 0.861$\\
    \hline
    \end{tabular}
    }
    \vspace{.05in}
    \caption{We ran Wilcoxon Rank-sum tests for all pairs of our four groups: Adaptive Recommendation~(A.R.), Non-adaptive Recommendation~(N.R.), Assessment-Only~(A.O.) and Random~(Rand.). The difference between adaptive recommendation and non-adaptive recommendation was statistically significant~($p=.035$).}   
    \label{tab:jrec_stats}
\end{table}

In order to measure engagement , we recorded the number of texts each user read before leaving. 131 randomly selected users used adaptive recommendation~(A.R.), 91 users used non-adaptive recommendation~(N.R.), 115 users used assessment-only~(A.O.) and 31 users used the random algorithm~(Rand.).\footnote{Users were assigned to these four conditions at a ratio of 3:3:3:1, respectively. Since the tool only recorded when a user responded to a text, the number of recorded users in each group differs somewhat from the expected ratio. This may be because some users quit before responding to the first problem.} Since our data was not normally distributed, we ran Wilcoxon Rank-sum tests for all pairs of the four groups~(Table~\ref{tab:jrec_stats}) . We observed that the median user in the adaptive recommendation group~($Median=13$) read 62.5\% more text than that in the non-adaptive recommendation group~($Median=8$), and the difference between these two groups was statistically significant~($p=.035$), which indicates that adaptive recommendation led users to read more texts than non-adaptive recommendation. Figure~\ref{fig:jrec_results} shows the proportions of users remaining after reading certain amounts of texts in the adaptive recommendation and the non-adaptive recommendation group. In addition, the median user in the assessment-only group read 12 texts, which was also significantly more than that in the non-adaptive recommendation group~($p=.022$). The median user in the random group read 8 texts and we did not find a statistically significant difference compared to the other three groups, possibly because the random group had too few users. Overall, our results demonstrate that incorporating adaptive assessment can significantly enhance learning material recommendation in online learning.

\section{Conclusion}

Recommending personalized learning materials in online language learning requires evaluation of the difficulty of learning materials and assessment of students' knowledge. Ideally, this would not require prior information about students or significant expert labor. To address this, we proposed a refined hierarchical knowledge structure to model vocabulary knowledge in authentic learning materials collected online. This model relaxes constraints on judgements of relative difficulty to ensure the knowledge structure is sufficiently dense. We also introduced a hybrid recommendation approach that balances assessment and recommendation in order to adapt to a student's prior knowledge. We evaluated these ideas in a Japanese learning text recommendation tool, and demonstrated that our adaptive recommendation approach engaged users for greater lengths of time than the non-adaptive version. In the future, we hope to incorporate other types of multimedia content and apply our model to other educational domains such as programming languages, mathematics, or even general knowledge. 


\section{Acknowledgements}
This material is based upon work supported by the National Science Foundation under Grant No. IIS-1657176.

%
%
%
\bibliographystyle{splncs04}
\bibliography{AIED2019}

\end{document}